\documentclass[authoryear]{elsarticle}
\usepackage{aas_macros}
\usepackage{amsfonts}
\usepackage[english]{babel}
\usepackage{natbib}
\usepackage{makeidx}
\usepackage{amsmath}
\usepackage{amssymb}
\usepackage{graphicx}

\begin{document}

\title{Spreading of ultrarelativistically expanding shell:\ an application
to GRBs}
\author{R. Ruffini}
\author{I.~A. Siutsou\corref{cor1}}
\ead{siutsou@icranet.org}
\author{G.~V. Vereshchagin}
\cortext[cor1]{Corresponding author}
\address{International Center of
Relativistic Astrophysics Network (ICRANet), Piazza
della Repubblica 10, 65122 Pescara, Italy.\smallskip\\
Dipartimento di Fisica, Universit\`a degli Studi di Roma ``Sapienza'', P.le
Aldo Moro 5, 00185 Roma, Italy.\\
}
%\date{}
\begin{abstract}
Optically thick energy dominated plasma created in the source of Gamma-Ray
Bursts (GRBs) expands radially with acceleration and forms a shell with
constant width measured in the laboratory frame. When strong Lorentz factor
gradients are present within the shell it is supposed to spread at sufficiently
large radii. There are two possible mechanisms of spreading: hydrodynamical and
thermal ones. We consider both mechanisms evaluating the amount of spreading
that occurs during expansion up to the moment when the expanding shell becomes
transparent for photons. We compute the hydrodynamical spreading of an
ultrarelativistically expanding shell. In the case of thermal spreading we
compute the velocity spread as a function of two parameters: comoving
temperature and bulk Lorentz factor of relativistic Maxwellian distribution.
Based on this result we determine the value of thermal spreading of
relativistically expanding shell. We found that thermal spreading is negligible
for typical GRB parameters. Instead hydrodynamical spreading appears to be
significant, with the shell width reaching $\sim10^{10}$~cm for total energy
$E=10^{54}$~erg and baryonic loading $B=10^{-2}$. Within the fireshell model
such spreading will result in the duration of Proper Gamma-Ray Bursts up to
several seconds.
\end{abstract}
\maketitle

\section{Introduction}

Optically thick pair plasma with baryon loading is assumed to power GRBs in
many models considered in the literature, see, e.g.,
\citet{1999PhR...314..575P}, \citet{2006RPPh...69.2259M},
\citet{2009AIPC.1132..199R}. Such plasma is self accelerated to large bulk
Lorentz factors due to initial energy dominance. Due to relativistic kinematics
in laboratory reference frame it forms a shell with width approximately equal
to its initial size. When the plasma becomes transparent to Compton scattering
a flash of radiation is emitted. In the literature such emission is often
associated with so-called precursors \citep{2005MNRAS.357..722L}. In the
fireshell model \citep{2009AIPC.1132..199R} this emission is called the Proper
GRB (P-GRB).

For typical baryonic loading parameters of observed GRBs $10^{-3}<B<10^{-2}$,
where $B=Mc^{2}/E_{0}$, $Mc^{2}$ is total baryonic rest mass energy, $E_{0}$ is
total energy released in the source of GRB, the fraction of energy emitted at
transparency can reach several percents of the total energy. Considering that
typical duration of long GRBs is of the order of hundreds of seconds, and P-GRB
typically lasts for less than few seconds \citep{2009AIPC.1132..199R}
luminosities of both events are comparable in magnitude.

Early treatments assuming the absence of baryons and thin shell approximation
\citep{2001A&A...368..377B}, and non instantaneous energy release
\citep{2005IJMPD..14..131R} produced estimations of P-GRBs duration $\sim
10^{-2}-10^{-1}$ sec for the progenitor mass range $10-10^{3}M_{\odot }$, much
larger than the naive estimation of the gravitational collapse time $\sim
GM/c^{3}\simeq 5\cdot 10^{-6}M/M_{\odot }$ sec, where $G$ is Newton's constant,
$M$ is black hole mass, $c$ is the speed of light. Observed durations of P-GRBs
\citep{2009AIPC.1132..199R} of most energetic GRBs is of the order of several
seconds. Main purpose of this paper is to outline a possibility of resolution
of this tension considering different mechanisms of spreading of
ultrarelativistically expanding shell.

The paper is organized as follows. In Sec. 2 the hydrodynamic phase of GRB is
discussed, hydrodynamical spreading mechanism is recalled, and results of its
application to GRBs are shown. In Sec. 3 the concept of thermal spreading is
introduced, evaluation of spreading for relativistic Maxwellian distribution
function is performed, and the results of thermal spreading computation are
applied to GRBs. Discussion of the spreading effects on the duration of P-GRB
and conclusions follow in the last section.

\section{Hydrodynamic phase and optical depth of GRBs}

Energy, entropy and baryonic number conservation laws for each differential
subshell of ultrarelativistically expanding shell imply
\citep{2012arXiv1205.3512R}
\begin{gather}
(p+\rho )\Gamma ^{2}r^{2}dr=\mathrm{const},  \\
\sigma \Gamma r^{2}dr=\mathrm{const}, \label{1} \\
n\Gamma r^{2}dr=\mathrm{const},
\end{gather}%
where $\rho $ and $p$ are respectively comoving energy density and pressure,
$\sigma$ and $n$ are respectively comoving entropy and comoving baryon number
density, $\Gamma $ is Lorentz factor, and $r$ is laboratory radial coordinate.
Assuming polytropic equation of state
\begin{equation}
p=\left( \gamma -1\right) \rho ,\quad \sigma =\rho ^{\gamma },
\end{equation}
where $\gamma $ is the thermal index, equations (1) can be solved for each
differential subshell in ultrarelativistic and nonrelativistic asymptotic
cases, respectively. Recall that expanding plasma consists of two components:
nonrelativistic baryons and ultrarelativistic electron-positron pairs and
photons. Hence, in analogy with cosmology, see e.g.
\citet{2012arXiv1205.3512R}, these two limiting cases are characterized by two
adiabatic indices, respectively: $\gamma=4/3$ when comoving temperature is
relativistic (energy dominated phase), and $\gamma \simeq 1$ when comoving
temperature is nonrelativistic (matter dominated phase). In the latter case the
pressure is usually neglected, see e.g. \citet{1993MNRAS.263..861P}. In what
follows we assume $p=0$ at the matter dominated phase. Therefore one has
\begin{equation}
\Gamma \propto r,\quad \rho \propto r^{-4},\quad n\propto r^{-3}
\end{equation}
during the energy dominated phase and
\begin{equation}\label{NRscalings}
\Gamma \simeq B^{-1}=\mathrm{const},\quad \rho \propto r^{-2},\quad n\propto
r^{-2}
\end{equation}
during the matter dominated phase. These scaling laws are different from the
ones in cosmology, see \citet{2012arXiv1205.3512R} for details and comparison.
The transition from energy to matter domination occurs at $R_{eq}\simeq
R_{0}/B$, where $R_{0}$ is initial size of the energy dominated region.

In \cite{2013ApJ...772...11R} we computed the optical depth and dynamics of
P-GRB for the case when $\Gamma$ is constant across the shell. General
expression for the optical depth of the shell is
\begin{equation}\label{tau}
    \tau=\int_R^{R^*}\sigma_T n\Gamma(1-\beta)dr
\end{equation}
where $\sigma _{T}$\ is Thomson's cross section and $R^*$ is the radius at
which light ray emitted from the inner boundary of the shell at radius $R$
crosses its outer boundary. For large baryonic loading with
\begin{equation}
B\gg 1\times 10^{-3}\hfill E_{54}^{-1/5}R_{8}^{2/5},  \label{condition}
\end{equation}
where $R_{0}=R_{8} 10^{8}$ cm, $E_{0}=E_{54} 10^{54}$ erg, optical depth of the
shell is
\begin{equation}\label{tauthin}
    \tau=\tau_0\left(\frac{R_0}{R}\right)^2,
\end{equation}
where $\tau_0=\sigma_T n_0 R_0$, $n_0$ is given by equation (\ref{n0}) with
$r_0=R_0$. Equation (\ref{tauthin}) correspond to the photon thin asymptotics
\citep{2013ApJ...772...11R} of equation (\ref{tau}) defined by the condition
$R^*-R=2\Gamma^2 R_0\ll R$.

The observed duration of P-GRB is determined in this asymptotics by the
process of radiative diffusion \citep{2013ApJ...772...11R}
\begin{equation}\label{tathin}
    \Delta t_a=\frac{t_{D}}{2\Gamma^2}
     \simeq 0.12 E_{54}^{1/3}B_{-2}^{5/3} R_8^{1/3}\text{ s},
\end{equation}
where $B=B_{-2}10^{-2}$ and time of diffusion $t_D$ was found by solution of
radiative transfer equation.

All these results are derived assuming constant Lorentz factor across the
shell. Hence the width of the shell remains constant during its expansion. This
fact has been used within the fireshell model and termed as the ``constant
thickness approximation'', provided that the baryon loading is not too heavy
$B\leq10^{-2}$ \citep{2000A&A...359..855R}. Hydrodynamic analytical
(\citet{1990ApJ...365L..55S}, \citet{1995PhRvD..52.4380B}) and numerical
(\citet{1993MNRAS.263..861P}, \citet{1993ApJ...415..181M}) calculations show
that Lorentz factor gradient is developing during shell expansion that leads to
spreading of the shell at sufficiently large radii at the matter dominated
phase.

%We consider the effect of hydrodynamical spreading on the duration of light
%signal from transparency of relativistic plasma \citep{2009AIPC.1132..199R}. We
%show that hydrodynamic mechanism may lead to sufficient spreading, which can
%account for observed duration of P-GRBs reaching up to a second.

At this phase of expansion each differential subshell is expanding with almost
constant speed $v=\beta c\simeq c(1-1/2\Gamma^{2})$, so the spreading of the
shell is determined by the radial dependence of the Lorentz factor $\Gamma(r)$.
In a variable shell there can be regions with $\Gamma(r)$ decreasing with
radius and $\Gamma(r)$ increasing with radius. At sufficiently large radii only
the regions with increasing $\Gamma$ contribute to the spreading of the shell.
From equations of motion of external and internal boundaries of this region we
obtain \citep{1993MNRAS.263..861P} the thickness of the region as function of
radial position of the region
\begin{equation}\label{lR}
    l(R)=R_0+\frac{R}{2}\left(\frac1{\Gamma_i^2}-\frac1{\Gamma_e^2}\right),
\end{equation}
where $\Gamma_e$ and $\Gamma_i$ are Lorentz factors at external and internal
boundaries, $R_0$ is the width of the region at small $R$, $R$ being the radial
position of inner boundary. Let us consider such a region in two limiting
cases:
\begin{itemize}
    \item[(a)]when relative Lorentz factor difference is strong,\\
        $\Delta\Gamma=\Gamma_e-\Gamma_i\gtrsim\Gamma_i$;
    \item[(b)] when relative Lorentz factor difference is weak,\\
        $\Delta\Gamma= \Gamma_e-\Gamma_i \ll \Gamma_i$.
\end{itemize}

In the case (a) the second term in parenthesis in equation~(\ref{lR}) can be
neglected, and we obtain that the spreading becomes efficient at
$R>R_b=2\Gamma_i^2 R_0$, see \cite{1993ApJ...415..181M} and
\cite{1993MNRAS.263..861P}. In the case (b) we find the corresponding critical
radius of hydrodynamical spreading $R_b=(\Gamma_i/\Delta\Gamma)\Gamma_i^2 l\gg
\Gamma_i^2 R_0$. From equation (\ref{lR}) one can see that in both cases for $R
\gg R_b$ width of the shell is increasing linearly with radius $l(R)\simeq
(\Delta\Gamma/\Gamma_i) R/\Gamma_i^2$.

The discussion above corresponds to the case (b) since for weak Lorentz factor
difference $R_b\gg R_{tr}$. In what follows we focus on the case (a) and derive
corresponding relations assuming strong relative Lorentz factor difference
across the shell. Let us take an element of fluid with constant number of
particles $dN$ in the part of the shell with gradient of $\Gamma$. Internal
boundary of the element is moving with velocity $v$, and external one is moving
with velocity $v+dv=v+\frac{dv}{dr}dr$, where $dr$ is the differential
thickness at some fixed laboratory time $t=0$ and derivative ${dv}/{dr}$ is
taken at the same laboratory time. Then at time $t$ the width of the element is
$dl=dr+t dv$, its radial position is $R(t)=r_0+v t$, where $r_0$ ia initial
radial position of the element, and corresponding laboratory density is
\begin{equation}\label{nspr}
    n_{l}=\frac{dN}{dV}=\frac{dN}{4\pi R^2 \left(1+ t \frac{dv}{dr}\right) dr}
        =n_0\frac{r_0^2}{R^2\left(1+ t \frac{dv}{dr}\right)},
\end{equation}
where
\begin{equation}\label{n0}
    n_0=\frac{dN}{dV_0}=\frac{dN}{4\pi r_0^2 dr}.
\end{equation}
At large enough $t$ using $R\simeq c t$ we have in contrast with
(\ref{NRscalings})
\begin{gather}\label{scalingsFARAWAY}
    \Gamma \simeq \mathrm{const}, \quad \rho\propto r^{-4}, \quad n\propto r^{-3},\\
    R\gg R_{b}=%\left(\frac{dv}{dr_0}\right)^{-1}c
    \frac{1}{\Gamma^3}\left(\frac{d\Gamma}{dr}\right)^{-1}. \nonumber
\end{gather}

In order to compute the integral (\ref{tau}) we need to find the expression for
baryonic number density along the light ray. Taking into account hydrodynamical
spreading (\ref{lR}) we obtain
\begin{equation}\label{nspralongtheray}
    n=\frac{n_0}{\Gamma}\left(\frac{R_0}{r}\right)^2
        \frac{1}{1+\frac{2 r}{\Gamma}\frac{d\Gamma}{dr}},
\end{equation}
that is exact in ultrarelativistic limit. Notice the difference between
equations~(\ref{nspralongtheray}) and (\ref{nspr}): in the former case
$d\Gamma/dr$ is computed along the light ray, while in the latter case $dv/dr$
is computed along the radial coordinate at fixed laboratory time. The
expression (\ref{nspralongtheray}) reduces to equation~(\ref{NRscalings}) when
$d\Gamma/dr=0$. Instead when the second term in the denominator of the
expression~(\ref{nspralongtheray}) dominates, namely when $r\gg
\Gamma(d\Gamma/dr)^{-1}$, density radial dependence coincides with the one
given by relations~(\ref{scalingsFARAWAY}).

%in this case light ray, emitted from the inner boundary at radius $R$, leave
%the shell at radius much larger than $R$. Along the light ray
An estimate for $d\Gamma/dr$ can be given for strong relative Lorentz factor
difference $\Delta\Gamma\sim\Gamma$ in the shell
\begin{equation}\label{dGdr}
    \frac{d\Gamma}{dr}\sim\frac{\Delta\Gamma}{\Delta r}\sim
        \frac{\Gamma}{2\Gamma^2 R_0}=\frac{1}{2\Gamma R_0},
\end{equation}
where $\Delta r\sim 2\Gamma^2 R_0$ is the distance inside the shell along the
light ray. Numerical results from \citet{1993MNRAS.263..861P},
\citet{1993ApJ...415..181M} and analytical ones from
\citet{1995PhRvD..52.4380B} support this estimate.

Integrating expression~(\ref{dGdr}) we obtain Lorentz factor dependence on
radial coordinate along the light ray
\begin{equation}%\label{}
    \Gamma(r)\sim\sqrt{\frac{r-R}{R_0}}.
\end{equation}
%Then from Eq.~(\ref{tau}) we have the integral
%\begin{equation}\label{tauspr}
%    \tau=\tau_0\int_{R}^{R^*} \left(\frac{R_0}{r}\right)^2
%        \frac{1}{1+\frac{2 r}{\Gamma}\frac{d\Gamma}{dr}}
%        \frac{1}{2\Gamma^2} dr.
%\end{equation}
Since we are interested in the asymptotics when the hydrodynamical spreading is
essential, we can assume in the integral (\ref{tau}) $r\gg R$ and $R^*\gg R$.
Under these conditions the optical depth is
\begin{equation}\label{tauspreading}
    \tau=\frac{\tau_0}{8}\left(\frac{R_0}{R}\right)^2.
\end{equation}
This result coincides with equation~(\ref{tauthin}) up to a numerical factor.
However its physical meaning is different. It represents photon thick
asymptotics of equation (\ref{tau}), since $R^*\gg R$
\citep{2013ApJ...772...11R}.

%and transparency radius is
%\begin{equation}
%R_{tr}\simeq\left( \frac{1}{32\pi }\frac{\sigma _{T}}{m_{p}c^{2}}BE_{0}\right)
%^{1/2}= 2\times 10^{14}B_{-2}^{1/2}E_{54}^{1/2}\text{ cm}.  \label{Rtrspr}
%\end{equation}
Transparency radius is defined by equating (\ref{tauspreading}) to unity and is
given by
\begin{equation}
R_{tr}\simeq\left( \frac{\sigma _{T} BE_{0}}{32\pi\,m_{p}c^{2}}\right)
^{1/2}= 2\times 10^{14}B_{-2}^{1/2}E_{54}^{1/2}\text{ cm},  \label{Rtr}
\end{equation}
where $m_{p}$ is proton mass. At the radius of transparency the width of the
shell (\ref{lR}) spreads up to
\begin{equation}
\Delta l_{hydr}\simeq
10^{10}\,B_{-2}^{1/2}\,E_{54}^{1/2}\,\Gamma_2^{-2}\text{ cm},  \label{hydrspr}
\end{equation}
where $\Gamma_i=100\Gamma_2$.

In this case radiative diffusion is irrelevant and duration of P-GRB is
then given by the time of arrival of photons emitted from the shell all the way
up to $R_{tr}$
\begin{equation}\label{taspr}
    \Delta t_a\simeq\frac{R_{tr}}{2\Gamma_i^2 c}=
     0.3\,B_{-2}^{1/2}\,E_{54}^{1/2}\,\Gamma_{2}^{-2}\text{ s}.
\end{equation}
In contrast to equation (\ref{tathin}) in the case of strong Lorentz factor
difference, namely with $\Gamma_i<(2B)^{-1}$, this duration can be of the order
of several seconds with agreement with observations.

%Note that time duration of P-GRB emitted as expanding shell becomes transparent
%to radiation is determined by two effects: the gradual transparency of the
%shell for photons emitted from different layers \citep{2011arXiv1110.0407R},
%and the delay of the light signal coming from different angles
%\citep{1986ApJ...308L..47G, 1996ApJ...473..998F}. In photon thick case
%transparency for all the layers of the shell is achieved at the same laboratory
%radius, so that the duration of the light signal from transparency is given by
%$\Delta t_a=l(R_{tr})/c$, since angular effect is negligible. In the photon
%thin case both gradual transparency of the shell and angular effect give
%$\Delta t_a=R_{tr}/(2\Gamma^2 c)$.

%However, in the considered case of thin shell due to hydrodynamical spreading
%both of these scales are close to each other and
%\begin{equation}%\label{}
%    \Delta t_a\simeq 0.3\,B_{-2}^{1/2}\,E_{54}^{1/2}\,\Gamma_{2}^{-2}\text{ s}.
%\end{equation}

\section{Thermal spreading}

We now determine the velocity spread of particles as a function of comoving
temperature $T$ and bulk Lorentz factor $\Gamma$ for relativistic Maxwellian
distribution. Based on this result we compute the value of thermal spreading of
expanding shell.

%We assume that each layer of the expanding shell is in local thermodynamical
%equilibrium. It is a reasonable assumption for the hydrodynamic phase of
%expansion due to large optical depth of the shell. Then
The distribution of particles in the momentum space in the laboratory frame
is a Lorentz-boosted Maxwellian
\begin{multline}  \label{Maxwellboost}
f(p_{x},p_{y},p_{z})=A\exp\Biggl(-\frac{1}{mc\theta}\Bigl[m^{2}c^{2}+{p_{y}}%
^{2}+{p_{z}}^{2} \\
+\left( \Gamma{p_{x}}-\sqrt{(\Gamma^{2}-1)(m^{2}c^{2}+{\mathbf{p}}^{2})}%
\right) ^{2} \Bigr]^{1/2}\Biggr) ,
\end{multline}
where we assumed that the relative motion of the frames is along their $x$%
-axes and $\theta=kT/mc^2$ is dimensionless comoving temperature.

Velocity dispersion in the $x$-direction is
\begin{equation}
D(v_{x})=M(v_{x}^{2})-M^{2}(v_{x}),   \label{dispersion}
\end{equation}
where $M(\chi)$ denotes the average value of $\chi$, which is defined by the
convolution with the distribution function (\ref{Maxwellboost})
\begin{equation}
M(\chi)=\frac{\displaystyle\int d^3\mathbf{p}\, \chi(\mathbf{p})f(\mathbf{p})%
}{\displaystyle\int d^3\mathbf{p} \,f(\mathbf{p})}.
\end{equation}
%In what follows we use the dimensionless velocity $\beta=v/c$.
The above written integrals cannot be computed analytically, but their
numerical approximations can be found.

Numerical issues in the velocity dispersion calculations by (\ref{dispersion})
arise from the fact that for high $\Gamma$ we need to subtract two numbers
$M(v^{2})$ and $M^{2}(v)$ which are very close to each other and to $c^2$. This
leads to substantial reduction of accuracy. A different formula for
dispersion%
\begin{equation}
D(v_{x})=M([v_{x}-M(v_{x})]^{2})   \label{dispersiondef}
\end{equation}
proves to be more convenient. The spread of particle velocities is then $%
(\Delta v)_{therm}=\sqrt{D(v)}$.

For nonrelativistic comoving temperatures the correct asymptotics is
\begin{equation}  \label{asymptotics}
\left(\frac{\Delta v}{c}\right)_{\theta\ll1}=\Gamma^{-2}\theta^{1/2}.
\end{equation}
The case of ultrarelativistic comoving temperature ($\theta\gg1$) is more
interesting. Starting close to the maximal value $1/\sqrt{2}$, the velocity
spread for $10\lesssim\Gamma \lesssim\theta$ reaches approximately
\begin{equation}
\left(\frac{\Delta v}{c}\right)_{10\lesssim\Gamma\lesssim\theta}
\simeq\Gamma^{-3/2},  \label{urmediumasymptotics}
\end{equation}
which means that the dispersion is independent on the temperature. For
$\Gamma\gg\theta$ the asymptotics (\ref{asymptotics}) is restored just up to a
multiplier close to unity
\begin{equation}
\left(\frac{\Delta v}{c}\right)_{1\ll\theta\ll\Gamma}
\simeq1.16\,\Gamma^{-2}\theta^{1/2}.   \label{urhighasymptotics}
\end{equation}

Our results suggest that (\ref{urmediumasymptotics}) gives absolute upper
limit for the velocity spread, and temperature dependence of (\ref%
{asymptotics}) and (\ref{urhighasymptotics}) reduce the spread even further.

%\section{Implications of thermal spreading for GRBs}

Initial temperature of the plasma formed in the source of GRB can be
estimated from its initial size $R_{0} \sim10^{8}$ cm and total energy
released
\begin{equation*}
10^{48}\text{ erg}<E_{0}<10^{55}\text{ erg}.
\end{equation*}
Assuming that the temperature is determined by $e^{+}e^{-}$\ pairs only
\begin{equation}  \label{temp}
\frac{E_{0}}{V_{0}}=\frac{3E_{0}}{4\pi R_{0}^{3}}=aT_{0}^{4}
\end{equation}
we get for initial temperature
\begin{equation}
0.40<\frac{kT_{0}}{m_{e}c^{2}}<22.
\end{equation}
At radiation dominated phase the comoving temperature of the plasma decreases
as $T\simeq T_{0}R_{0}/r$, at matter dominated phase as $T\simeq
T_{0}B^{1/3}(R_{0}/r)^{2/3}$. Now we compute the thermal spreading at both
phases.

For the first phase the reasonable approximation of Lorentz factor is
\begin{equation*}
\Gamma(t)\simeq\sqrt{1+\left( \frac{ct}{R_{0}}\right) ^{2}}.
\end{equation*}
Due to the nature of Lorentz transformations in constantly accelerated frame
%(as in the case of hyperbolic motion),
the final spreading of the shell $\Delta l_{1}=\int_{0}^{t}\Delta v\,dt$
appears to be finite even if we extend this phase infinitely in time, and the
main part of the spreading is connected with initial part of motion with
relatively small $\Gamma$.

While the temperature of protons in the source of GRB is nonrelativistic
$\theta\ll1$, velocity spread is given by (\ref{asymptotics}) which in energy
dominated phase leads to the spreading
\begin{equation}  \label{nr1}
\frac{\Delta l_{1}}{R_{0}}\lesssim2.2\sqrt{\frac{kT_{0}}{m_pc^{2}}}
=0.18\,E_{54}^{1/8}\,R_{8}^{-3/8}.
\end{equation}

In the matter dominated phase the additional spreading of the shell is
\begin{equation}
\frac{\Delta l_{2}}{R_{0}}%=\int_{t_{1}}^{t_{tr}}\Delta v(t)dt=\int_{t_{1}}^{t_{tr}}
%c\Gamma^{-2}\sqrt{\frac{kT_{0}R_{0}}{mc^{2}ct}}dt
\simeq3B^{7/3}\sqrt{\frac{kT_{0}}{m_pc^{2}}}\,\left(\frac{R_{tr}}{R_0}\right)^{1/3}%=\\
=6.7\cdot10^{-4}\,E_{54}^{7/24}\,R_{8}^{-17/24}\,B_{-2}^{5/2}, \label{nr2}
\end{equation}
when $R_{tr}\gg R_{0}$. Comparing to the hydrodynamical spreading for
reasonable GRB parameters the spreading coming from both (\ref{nr1}) and
(\ref{nr2}) is negligible.

Note that velocity dispersion in any case does not exceed the value given by
equation~(\ref{urmediumasymptotics}) with $\theta \gg 1$
\begin{equation}
\frac{\Delta l_{1}}{R_{0}}=\frac{1}{R_{0}}\int_{0}^{t_{1}}\Delta
v(t)dt\lesssim \frac{1}{R_{0}}\int_{0}^{\infty }c\Gamma (t)^{-3/2}dt\simeq
2.6,  \label{ur1}
\end{equation}
which gives an absolute maximum of the thermal spreading on the energy
dominated phase.

\section{Conclusion}

In this paper we considered two mechanisms of spreading of relativistically
expanding plasma shell. We also discuss their implications for the duration of
electromagnetic signal from transparency of the plasma.

Firstly, following the proposal of \citet{1993MNRAS.263..861P} hydrodynamical
spreading of relativistically expanding shell is estimated. Secondly, thermal
spreading is considered. Assuming relativistic Maxwellian distribution function
we determined the velocity dispersion depending on temperature and the Lorentz
factor of the bulk motion.

We then applied these results to GRBs within the framework of the fireshell
model. It is shown that thermal spreading provides negligible spreading for
typical parameters of GRBs. Instead, hydrodynamical spreading results in the
increase of the duration of P-GRB. For nonspreading shells characterized by
almost constant Lorentz factor distribution within the shell the duration of
P-GRB is determined by the time of diffusion, see equation (\ref{tathin}). If
strong Lorentz factor difference is present within the shell, hydrodynamical
spreading prevents occurrence of photon thin asymptotics and leads to duration
of P-GRB given by equation (\ref{taspr}).

Our results imply that for high enough baryon loading and energy of the burst
the duration of the P-GRB is not determined by the initial size of the plasma
$R_{0}$, but by the value of the hydrodynamical spreading, reaching up to
several seconds.

%\bibliographystyle{elsarticle-harv}
%\bibliography{tapp}

\end{document}